\documentclass[prb,twocolumn,showpacs,preprintnumbers,amsmath,amssymb,floatfix]{revtex4}

\usepackage{graphicx}
\usepackage{dcolumn}
\usepackage{bm}

\begin{document}
\title{Electron-phonon superconductivity in
non-centrosymmetric LaNiC$_2$: first principles calculations}
\author{Alaska Subedi}
\affiliation{Department of Physics and Astronomy, University of Tennessee,
  Knoxville, Tennessee 37996, USA}
\affiliation{Materials Science and Technology Division, Oak Ridge National
  Laboratory, Oak Ridge, Tennessee 37831-6114, USA}
\author{David J. Singh}
\affiliation{Materials Science and Technology Division, Oak Ridge National
  Laboratory, Oak Ridge, Tennessee 37831-6114, USA}

\begin{abstract}
 We report first principles calculations of the electronic structure
and electron-phonon coupling in the non-centrosymmetric
superconductor LaNiC$_2$. These show that the material is a
 conventional electron-phonon superconductor with intermediate coupling.
There are large contributions to the coupling by two
  low frequency C non-bond-stretching modes, one of which has strong Kohn
  anomalies. Since LaNiC$_2$ lacks inversion symmetry, the pairing is of
  dominant $s$-wave type with some mixture of $p$-wave character. This will
  give exponential decay in the specific heat.
\end{abstract}

\pacs{74.70.Dd,74.25.Jb,74.25.Kc}

\maketitle

LaNiC$_2$ is a member of a family of ternary nickel carbides $R$NiC$_2$ (where
$R$ is a rare earth or Y) first reported by Bodak and Marusin.\cite{boda79}
These compounds form in the base centered orthorhombic CeNiC$_2$-type
structure (space group $Amm2$) that lacks inversion symmetry. Studies of
magnetic properties show that the $R$NiC$_2$ phases exhibit a variety of
magnetic ordering for various $R$.\cite{kots89, onod98}
Although, Ni is often associated with magnetism, e.g. in oxides and
in intermetallics, the magnetism in these
compounds is due to 4$f$ electrons of $R$ with almost no magnetic contribution
from Ni. Therefore, it is perhaps
not surprising that LaNiC$_2$ does not show magnetic
ordering.
Interestingly,
it is instead superconducting at $T_c = 2.7$ K.\cite{lee96,pech98}

Lee \textit{et al.}\ found non-exponential decay of specific heat below $T_c$
and based on this
argued that LaNiC$_2$ is an unconventional superconductor.\cite{lee96}
Since pure Ni is a ferromagnet and Ni compounds are often magnetic,
one might suppose LaNiC$_2$ to be near magnetism
and therefore that a pairing mechanism involving spin fluctuations is
operative. Spin fluctuations are not pairing for order parameters
that do not change sign on the Fermi surface (i.e. conventional $s$-wave).
However, Pecharsky
\textit{et al.} observed the usual exponential decay of specific heat below
$T_c$, consistent with the conventional BCS superconductivity.\cite{pech98}
One avenue to address the inconsistency of these two experiments is by
determining the symmetry that is broken in the superconducting system. Gauge
symmetry is broken in a conventional BCS superconductor,\cite{bard57} while
other symmetries are broken in unconventional ones.\cite{sigr91}
Related to this,
Hillier \textit{et al.}\ recently reported their results of muon spin
relaxation measurements on LaNiC$_2$ that indicate
that time reversal symmetry is
broken in the superconducting state.\cite{hill09}
Such time reversal symmetry breaking requires a superconducting state with
triplet character, and does not occur in singlet superconductors.
However, this analysis is complicated by the non-centrosymmetric lattice
structure of LaNiC$_2$.

There are a number of
other examples of non-centrosymmetric superconductors including for
example
CePtSi$_3$, CeIrSi$_3$, CeRh$_3$Si, UIr, Li$_2$Pd$_x$B, Re$_3$W and Y$_2$C$_3$.
\cite{bauer,kimura,sugitani,togano,akazawa,hulm,amano}
These include both non-phonon mediated heavy Fermion metals and
electron phonon superconductors, which behave very much like conventional
centrosymmetric superconductors.
Since parity is not a good quantum number in non-centrosymmetric
systems, it necessitates the modification of scheme for classification of
Cooper pairs. For example, Yanase and Sigrist \cite{yana07}
argue that Cooper pairs have a
state that has dominant $p$-wave with some mixture of $s$-wave symmetry in
CePt$_3$Si.
Considering the $T_c$ of LaNiC$_2$, either case is possible;
it could be
a conventional electron-phonon superconductor in analogy with
Y$_2$C$_3$ (refs. \onlinecite{amano,singh-mazin}) and
Re$_3$W (ref. \onlinecite{zuev}),
in which case it would have
a Cooper pair state that has dominant $s$-wave but some mixture of
$p$-wave symmetry or alternatively
it could be an unconventional superconductor in analogy with
the non-centrosymmetric heavy Fermions such
as CePt$_3$Si (refs. \onlinecite{bauer,mukuda}), or near ferromagnetic UIr,
in which case it would have
dominant $p$ or $d$ wave character, which would allow a
simpler explanation of non-exponential behavior.

Here, we report the results of first principles calculations of
electronic structure, phonon dispersion and electron-phonon coupling.
We find that the main
contributions to the electronic
structure from Ni $3d$ orbitals are away from the
Fermi energy and consequently that
LaNiC$_2$ is non-magnetic and rather far from magnetic instabilities
that might lead to strong spin fluctuations thereby leading to an
unconventional superconducting state.
On the other hand, we obtain a value of electron-phonon
coupling constant $\lambda \sim 0.52$ with logarithmically averaged frequency
$\omega_{\textrm{ln}} \sim 207$ cm$^{-1}$. Using simplified Allen-Dynes
formula, we obtain $T_c \sim 3.0$ K that suggests LaNiC$_2$ is an
intermediately coupled electron-phonon superconductor with dominant pairing
that has $s$-wave but also some mixture of $p$-wave symmetry.

LaNiC$_2$ forms in a base centered orthorhombic structure with La on site
$(0.5,u,1-u)$, Ni on site $(0,0,0)$ and C on site $(0,v,w)$. In
this compound, the
La atoms form trigonal prisms, which are alternately filled by
Ni and C dimers, hence breaking the inversion symmetry. Another characteristic
of this structure is the short bond length of C dimers, which indicates the
existence of very stiff bond stretching modes. If these modes were
responsible for superconductivity, it would result in a moderately high
electron-phonon coupling and a large logarithmically averaged phonon
frequency. This would mean a very high prefactor in the McMillan equation for
$T_c$. Previous tight-binding\cite{lee89} and density functional
\cite{singh-mazin,guld97}
studies of other metals with C dimers show that the electronic structure
near the Fermi level could
have substantial contributions from
states with C-C antibonding character. If this were the case
for LaNiC$_2$, there
may be a strong dependence of $T_c$ on the occupation
of the
antibonding states (which could be controlled by doping) and perhaps also a
high $T_c$ in optimally doped samples. 

Our electronic structure calculations were performed within the local density
approximation (LDA) using the general potential linearized augmented planewave
(LAPW) method as
implemented in our in-house code.\cite{singbk}
We used LAPW spheres of radius 2.0$a_0$ for La and
Ni and 1.25$a_0$ for C.
We repeated some calculations with the WIEN2k code \cite{wien}
as a test and found no significant differences.
We used the
experimentally reported lattice parameters ($a = 3.959$ \AA,
$b = 4.564$ \AA, $c = 6.204$ \AA),\cite{boda79} but relaxed the internal
coordinates. We obtain for internal parameters, $u = 0.3886$, $v = 0.5411$,
and $w = 0.1606$. This yields the calculated C-C distance of 1.36 \AA, which
is in reasonable accord with the
reported experimental value \cite{boda79} of 1.41 \AA.

The phonon dispersions and electron-phonon coupling were calculated using
linear response as implemented in QUANTUM ESPRESSO code.\cite{qe} The linear
response calculations were also done using experimental lattice parameters,
using ultrasoft pseudopotentials within the generalized gradient approximation
(GGA) of Perdew \textit{et al.}\cite{perd96} An 8 $\times$ 8 $\times$ 8  grid
was used for the zone integration in the phonon calculations, while a more
dense 32 $\times$ 32 $\times$ 32 grid was used for the zone integration in the
electron-phonon coupling calculations. The basis set cutoff for the wave
functions was 40 Ry, while a 400 Ry cutoff was used for the charge
density. The internal coordinates were again relaxed and we obtained values
that agreed well with the values obtained from LDA calculations. 
The GGA and LDA electronic structures were very similar.

\begin{figure}[tbp]
  \includegraphics[scale=0.3,angle=270]{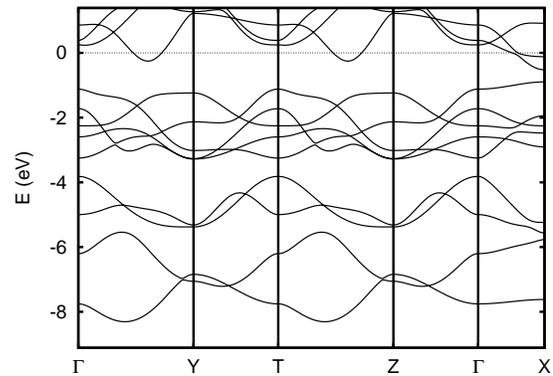}
  \caption{Calculated LDA band structure of LaNiC$_2$
    plotted along the path (0,0,0) $\rightarrow$ (0,$\frac{1}{2}$,0)
    $\rightarrow$ (0,$\frac{1}{2}$,$\frac{1}{2}$) $\rightarrow$
    (0,0,$\frac{1}{2}$) $\rightarrow$ (0,0,0) $\rightarrow$
    ($\frac{1}{2}$,0,0). }
  \label{fig:LaNiC2-bnd}
\end{figure}

\begin{figure}[tbp]
  \includegraphics[scale=0.3]{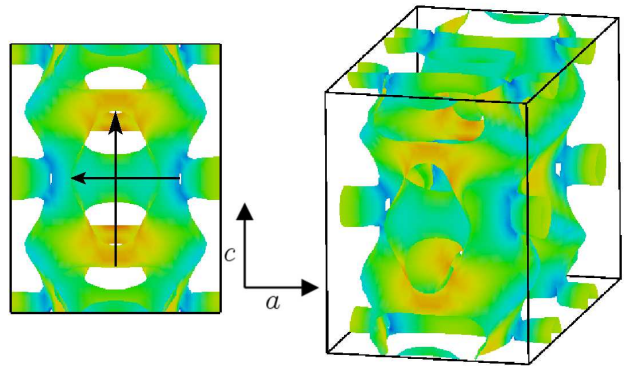}
  \caption{(Color online) Calculated LDA Fermi surface of
    LaNiC$_2$. The shading is by velocity. The arrows indicate nesting vectors
    that give rise to Kohn anomalies.}
  \label{fig:LaNiC2-fs}
\end{figure}

\begin{figure}[tbp]
  \includegraphics[scale=0.3,angle=270]{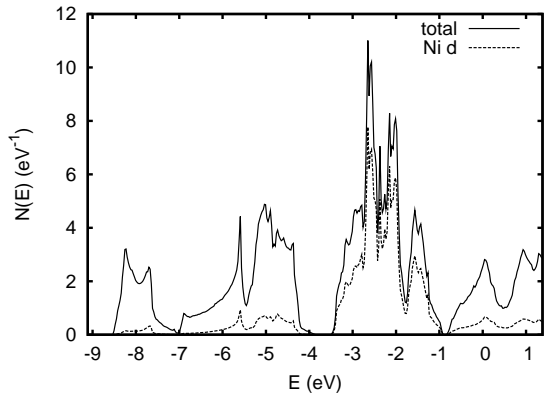}
  \caption{Calculated LDA density of states of
    LaNiC$_2$ on a per formula unit both spins basis. The projection is onto
  the Ni LAPW sphere.}
  \label{fig:LaNiC2-dos}
\end{figure}

\begin{figure}[tbp]
  \includegraphics[scale=0.3,angle=270]{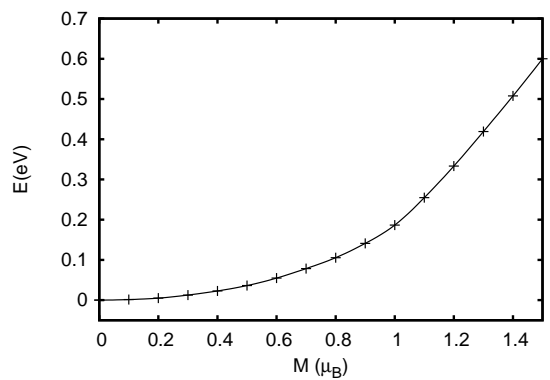}
  \caption{Fixed spin moment magnetic
 energies for LaNiC$_2$.}
  \label{fig:LaNiC2-ene-mom}
\end{figure}

The calculated band structure and Fermi surface of
LaNiC$_2$ is shown in Figs.~\ref{fig:LaNiC2-bnd} and \ref{fig:LaNiC2-fs},
respectively. These agree in large
scale features with recent calculations done by Laverock
\textit{et al.},\cite{lave09} but the details of the Fermi surface and
band dispersions differ significantly along some directions, yielding
differences in the Fermi surface.
These presumably reflect the use of non-full potential methods in Ref.
\onlinecite{lave09}.
The corresponding density of states is shown in
Fig.~\ref{fig:LaNiC2-dos}. The band structure shows a C $2s$ derived band
between -15.5 and -13.5 eV and another C $2s$ derived band between -8.5 and
-6.5 eV, relative to the Fermi energy $E_F$. This is followed by a manifold of
three bands with C $2p$ character, associated with C dimers, that extend from
-7.5 to -3.8 eV. This is followed by
a manifold of 5 bands of Ni $3d$ character
between -3.5 and -0.9 eV. These account for the majority of the Ni $d$
character in the density of states although there is some Ni $d$ contribution
near $E_F$, similar to the borocarbides (e.g. LuNi$_2$B$_2$C), which electron
phonon superconductors. \cite{pickett}
There are two bands that cross the Fermi level and
they have mixed Ni $3d$, La $5d$ and C $2p$ antibonding characters. This
electronic structure is consistent with strongly bonded C dimers embedded in a
metallic 
solid with states near $E_F$ having La, Ni and C antibonding
character.

The density of states at $E_F$ is $N(E_F) = 2.6$ eV$^{-1}$ on a per formula
unit both spin basis.
The calculated Fermi velocities are
$<v^2_{xx}>^{1/2}$=1.79x10$^5$ m/s,
$<v^2_{yy}>^{1/2}$=1.52x10$^5$ m/s, and
$<v^2_{zz}>^{1/2}$=1.48x10$^5$ m/s,
indicating only modest resistivity anisotropy unless there is a large
scattering anisotropy.
The density of states has a prominent peak
with a maximum very close to $E_F$. Therefore
the stoichiometric compound is expected to have the highest $T_c$
similar to Y$_2$C$_3$, although in the present case the peak is broader.
Doping away from stoichiometry will
move the $E_F$ away from the peak and is therefore expected
to lower the $T_c$. The moderately
high value of $N(E_F)$ might also
indicate magnetism if the states at the
Fermi level are of predominantly Ni $d$ character. However, as mentioned,
most of the Ni $d$ states are in a manifold of 5 bands that lie between
-3.5 and -0.9 eV relative to the Fermi level. Hence, the
high $N(E_F)$ of LaNiC$_2$ does not place it near
magnetism. This is supported by our fixed spin moment
calculations. Fig.~\ref{fig:LaNiC2-ene-mom} shows energies calculated for
different values of magnetic moment relative to the nonmagnetic case. 
As may be seen the energy scale is relatively high, which means that
the material is not near ferromagnetism.

\begin{figure}[tbp]
  \includegraphics[scale=0.4]{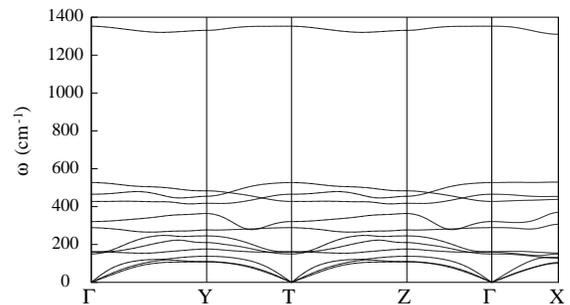}
  \caption{Calculated phonon dispersion curves of LaNiC$_2$ plotted along the
    path (0,0,0) $\rightarrow$ (0,$\frac{1}{2}$,0) $\rightarrow$
    (0,$\frac{1}{2}$,$\frac{1}{2}$) $\rightarrow$ (0,0,$\frac{1}{2}$)
    $\rightarrow$ (0,0,0) $\rightarrow$ ($\frac{1}{2}$,0,0).}
  \label{fig:LaNiC2-ph}
\end{figure}

\begin{figure}[tbp]
  \includegraphics[scale=0.3,angle=270]{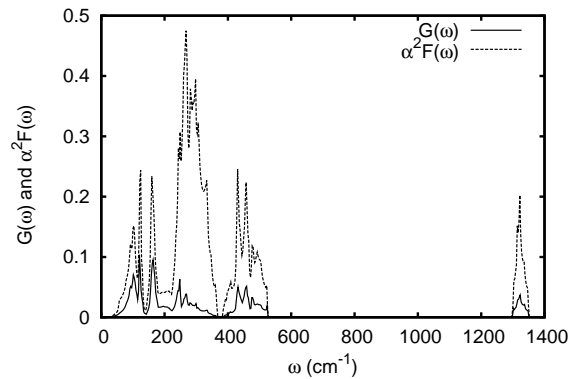}
  \caption{Calculated phonon density of states and electron-phonon spectral
    function $\alpha^2 F(\omega)$ for LaNiC$_2$.}
  \label{fig:LaNiC2-dos-a2F}
\end{figure}

The calculated phonon dispersions of LaNiC$_2$ are shown in
Fig.~\ref{fig:LaNiC2-ph}. The corresponding phonon density of states and
Eliashberg spectral function $\alpha^2 F(\omega)$ are shown in
Fig.~\ref{fig:LaNiC2-dos-a2F}. The phonon dispersions show a set of 11 bands
extending up to $\sim$ 530 cm$^{-1}$, separated by a gap of $\sim$ 775
cm$^{-1}$ from 1 high frequency band that lies between 1305 and 1355
cm$^{-1}$. The bands below 180 cm$^{-1}$ show contributions from La, Ni and C,
while the bands between 180 and 255 cm$^{-1}$ are derived mainly from motions
of Ni and C atoms. Above 255 cm$^{-1}$ the bands are dominated by motions of C
atoms. The highest frequency band is an almost pure C-C bond stretching mode,
confirming the picture of strongly bonded C dimer inferred from band
structure.

The eighth branch of the phonon dispersion, which has mainly C
non-bond-stretching character, shows Kohn anomalies at $k\sim(0,0.5,0.25)$,
$k\sim(0,0,0.25)$ and $k\sim(0.25,0,0)$. These can be seen in
Fig.~\ref{fig:LaNiC2-ph} along the Y--T, Z--$\Gamma$ and $\Gamma$--X lines.
The origin
of the Kohn anomalies can be seen in the Fermi surface, which shows nesting
with wavevectors $k\sim(0,0,0.25)$ and $k\sim(0.25,0,0)$ as indicated by two
arrows in Fig.~\ref{fig:LaNiC2-fs}. A comparison of phonon density of states
with the with the Eliashberg spectral function shows that the latter is
enhanced relative to the former between 220 and 400 cm$^{-1}$,
which is the region
where the seventh (which also has mainly C non-bond-stretching character) and
eighth branches are placed. We obtain a value of the electron-phonon coupling
$\lambda_{ep}=0.52$.
As there are 12 phonon branches, the average electron-phonon
coupling per branch is $\lambda_{avg}=0.04$. The contribution due to seventh
and eighth bands is large with values of $\lambda_7 = 0.13$ and $\lambda_8 =
0.09$, respectively. We obtain for the logarithmically averaged frequency
$\omega_{\ln} = 207$ cm$^{-1}$. Inserting these numbers into the simplified
Allen-Dynes formula,
\begin{equation}
  k_B T_c = {\frac{\hbar \omega_{\rm ln}}{1.2}} ~~ {\rm exp}
  \left \{
  - {\frac{1.04 (1 + \lambda_{ep})} {\lambda_{ep} - \mu^* (1 + 0.62
      \lambda_{ep})}}
  \right \} ,
\end{equation}
with $\mu^*=0.12$, we obtain $T_c \sim 3$ K, which is
in accord with the experimental value of $T_c =
2.7$ K.\cite{lee96,pech98}
While this close agreement may be partly fortuitous (note the
parameter $\mu^*$ in the Allen-Dynes formula; 0.12 is a typical value,
but values in the range 0.1 - 0.15 are reasonable),
it does indicate
that the superconductivity can be readily explained in an
electron-phonon framework.

To summarize, we report full potential calculations
of the electronic structure, phonon dispersions of LaNiC$_2$
and electron-phonon coupling in LaNiC$_2$.
We find that conventional electron-phonon coupling picture
readily describes the superconductivity of LaNiC$_2$.
With inversion center this would yield a pure $s$-wave state. However,
because of the non-centrosymmetric structure of LaNiC$_2$ a small
$p$ admixture will be present.
The Ni $d$ states are located mainly away from the
Fermi energy, and accordingly we do not find LaNiC$_2$
to be near magnetism.
Therefore,
we expect the contribution of spin fluctuations to the
coupling, which is pair breaking for $s$-wave, to be small.
The
phonon-mediated Cooper pairs are intermediately coupled with large
contributions made by two low frequency C non-bond-stretching modes, one of
which has Kohn anomalies.
Owing to the lack of inversion symmetry, the pairing
will have dominant $s$-wave character with some mixture of $p$-wave. This is
expected to give exponential decay of specific heat below $T_C$. In
particular, there will be no line nodes in the gap function.
Also, while the Cooper pairs also have some $p$-wave character,
a pure electron phonon mechanism will not yield breaking of
time reversal symmetry due to the removal of centrosymmetry.

We are grateful for helpful discussions with Lijun Zhang and support
from the Department of Energy, Division of Materials Sciences and Engineering.


\begin{references}
  \bibitem{boda79}
    O.I. Bodak and E.P. Marusin, Dopov. Akad. Nauk. Ukr. RSR, Ser. A:
    Fiz.-Mat. Tekh. Nauki \textbf{41}, 1048 (1979). 
    
  \bibitem{kots89} P. Kotsanidis and J.K. Yakinthos, Journal of the
    Less-Common Metals \textbf{152}, 287 (1989).

  \bibitem{onod98}
    H. Onodera, Y. Koshikawa, M. Kosaka, M. Ohashi, H. Yamauchi and
    Y. Yamaguchi, Journal of Magnetism and Magnetic Materials \textbf{182},
    161 (1998). 
    
  \bibitem{lee96}
    W.H. Lee, H.K. Zeng, Y.D. Yao, and Y.Y. Chen, Physica C \textbf{266}, 138
    (1996). 

  \bibitem{pech98}
    V.K. Pecharsky, L.L. Miller, and K.A. Gschneidner,
 Phys. Rev. B \textbf{58}, 497 (1998). 

  \bibitem{bard57}
    J. Bardeen, L.N. Cooper, and J.R. Schrieffer, Physical Review \textbf{108},
    1175 (1957).

  \bibitem{sigr91}
    M. Sigrist and K. Ueda, Rev. Mod. Phys. \textbf{63}, 239
    (1991). 

  \bibitem{hill09}
    A.D. Hillier, J. Quintanilla, and R. Cywinski, 
Phys. Rev. Lett. {\bf 102}, 117007 (2009).

\bibitem{bauer}
E. Bauer, G. Hilscher, H. Michor, C. Paul, E.W. Scheidt, A. Gribanov,
Y. Seropegin, H. Noel, M. Sigrist, and P. Rogl,
Phys. Rev. Lett. {\bf 92}, 027003 (2004).

\bibitem{kimura}
N. Kimura, K. Ito, K. Saitoh, Y. Umeda, H. Aoki, and T. Terashima,
Phys. Rev. Lett. {\bf 95}, 247004 (2005).

\bibitem{sugitani}
I. Sugitani, Y. Okuda, H. Shishido, T. Yamada, A. Thamizhavel,
E. Yamamoto, T.D. Matsuda, Y. Haga, T. Takeuchi, R. Settai, and Y. Onuki,
J. Phys. Soc. Jpn. {\bf 75}, 043703 (2006).

\bibitem{togano}
K. Togano, P. Badica, Y. Nakamori, S. Orimo, H. Takeya, and K. Hirata,
Phys. Rev. Lett. {\bf 93}, 247004 (2004).

\bibitem{akazawa}
T. Akazawa, H. Hidaka, T. Fujiwara, T.C. Kobayashi, E. Yamamoto, Y. Haga,
R. Settai, and Y. Onuki,
J. Phys. Condens. Matter {\bf 16}, L29 (2004).

\bibitem{hulm}
J.K. Hulm and R.D. Blaugher, 
J. Phys. Chem. Solids {\bf 19}, 134 (1961).

\bibitem{amano}
G. Amano, S. Akutagawa, T. Muranaka, Y. Zenitani, and J. Akimitsu,
J. Phys. Soc. Jpn. {\bf 73}, 530 (2004).

  \bibitem{yana07}
    Y. Yanase and M. Sigrist, J. Phys. Soc. Jpn. \textbf{76}, 043712 (2007).

\bibitem{singh-mazin}
D.J. Singh and I.I. Mazin, Phys. Rev. B {\bf 70}, 052504 (2004).

\bibitem{zuev}
Y.L. Zuev, V.A. Kuznetsova, R. Prozorov, M.D. Vannette, M.V. Lobanov,
D.K. Christen, and J.R. Thompson,
Phys. Rev. B {\bf 76}, 132508 (2007).

\bibitem{mukuda}
H. Mukuda, S. Nishide, A. Harada, K. Iwasaki, M. Yogi, M. Yashima,
Y. Kitaoka, M. Tsujino, T. Takeuchi, R. Settai, Y. Onuki, E. Bauer,
K.M. Itoh, and E.E. Haller,
J. Phys. Soc. Jpn. {\bf 78}, 014705 (2009).
    
  \bibitem{lee89}
    S. Lee, W. Jeitschko, and R.-D. Hoffmann, Inorg. Chem. \textbf{28}, 4094
    (1989). 
    
  \bibitem{guld97}
    Th. Gulden, R.W. Henn, O. Jepsen, R.K. Kremer, W. Schnelle, A. Simon, and
    C. Felser, Phys. Rev. B \textbf{56}, 9021 (1997).
    
  \bibitem{singbk}
    D.J. Singh and L. Nordstrom, \textit{Planewaves, Pseudopotentials and the
      LAPW Method, 2nd Ed.} (Springer, Berlin, 2006).

\bibitem{wien}
P. Blaha, K. Schwarz G.K.H. Madsen, D. Kvasnicka, and J.
Luitz, \textit{WIEN2K, An Augmented Plane Wave + Local Orbitals Program for
for Calculating Crystal Properties} (K. Schwarz, TU Wien,
Austria, 2001), ISBN 3-9501031-1-2


  \bibitem{qe}
    P. Giannozzi et al., http://www.quantum-espresso.org

  \bibitem{perd96}
    J.P. Perdew, K. Burke, and M. Ernzerhof, Phys. Rev. Lett. \textbf{77},
    3865 (1996).

  \bibitem{lave09}
    J. Laverock, T.D. Haynes, C. Utfeld and S.B. Dugdale, arXiv:0903.1814

\bibitem{pickett}
W.E. Pickett and D.J. Singh, Phys. Rev. Lett. {\bf 72}, 3702 (1994).

\end{references}
\end{document}